\documentclass[aps,prl,twocolumn,showpacs,amsmath,amssymb,groupedaddress]{revtex4}
\usepackage{graphicx}
\usepackage{dcolumn}
\usepackage{bm}
\usepackage{amsfonts}
\newcommand{\ket}[1]{\mbox{$\mid \! #1 \, \rangle$}}
\newcommand{\bra}[1]{\mbox{$\langle \, #1 \! \mid$}}

\newcommand{\eins}{\mbox{$1 \hspace{-1.0mm}  {\bf l}$}}

\newcommand{\WW}{\ensuremath{\mathcal{W}}}

\begin{document}
\title{Experimental test of rotational invariance and entanglement of
photonic six-qubit singlet state}
\author{Magnus R\aa dmark$^{1}$, Marek \.{Zukowski}$^{2}$ and Mohamed
Bourennane$^{1}$}
\affiliation{ $^{1}$Physics Department, Stockholm University, SE-10691 Stockholm, Sweden\\
$^{2}$Institute for Theoretical Physics and Astrophysics, Uniwersytet
Gda\'{n}ski, PL-80-952 Gda\'{n}sk, Poland}
\date{\today}
\begin{abstract}
We experimentally test invariance properties of the
six-photonic-qubits generalization of a singlet state. Our results
clearly corroborate with theory. The invariance properties are
useful to beat some types of decoherence. We also experimentally
detect entanglement in the state using an appropriate witness
observable, composed of local observables. Our results clearly
indicate that the tested setup, which is very stable, is a good
candidate for realization of various multi-party quantum
communication protocols. The estimated fidelity of the produced
state is very high: it reveals very strong EPR six-photon
correlations of error rate below $6\%$.
\end{abstract}
\maketitle
%

It is well known that quantum information processing  relies on
preparation, manipulation,  and detection of  superpositions of
quantum states. Superpositions, however, are very fragile and are
easily destroyed by the decoherence processes due to unwanted
couplings with the environment \cite{Z91}. Such uncontrollable
influences cause noise in the communication, or errors in the
outcome of a computation. Several strategies have been devised to
cope with decoherence. For instance, if the qubit-environment
interaction, no matter how strong, exhibits some symmetry, then
there exist quantum states which are invariant under this
interaction. These states are called decoherence-free (DF) states,
and allow to protect quantum information \cite{ZR97a,LCW98}. This
situation occurs for instance when the spatial (temporal) separation
between the carriers of the qubits is small relative to the
correlation length (time) of the environment. Experimental efforts
investigating features of DF systems have been carried out to
demonstrate the properties of a specific two-qubit DF state
\cite{KBAW00},  and the existence of three-qubit noiseless
subsystems \cite{VFPKLC01}. Bourennane {\it et al.} \cite{BEGKCW04}
produced the four-photon polarization-entangled state which is a
generalization of the singlet, $\ket{\Psi_{4}^{-}}$, and
demonstrated its invariance under general collective noise and
experimentally showed the immunity of a qubit encoded in this state.

To encode an arbitrary two-qubit state Cabello has theoretically
constructed  DF states formed by six qubits. One of these states is
$\ket{\Psi_{6}^{-}}$ \cite{C06}. It is invariant under
transformations which consist of identical unitary transformations
of each individual constituent\cite{ZR97a}:
\begin{equation}
U^{\otimes 6}\ket{\Psi_{6}^{-}} = \ket{\Psi_{6}^{-}},
\end{equation}
where $U^{\otimes 6} = U\otimes...\otimes U$ denotes the tensor
product of six identical unitary operators $U$. Besides protecting
against collective noise, the DF states are useful for communication
of quantum information between two observers who do not share a
common reference frame \cite{BRS03}. In such a scenario, any
realignment of the receiver's reference frame corresponds to an
application of the same transformation to each of the qubits which
were sent.
The states  $\ket{\Psi_{6}^{-}}$ can also be used for secure quantum
multiparty cryptographic protocols such as the  six-party secret
sharing protocol \cite{HBB99,GKBW07}.


Recently multiphoton interferometry based on parametric down
conversion reached the stage at which one can observe genuine
six-photon interference. The experiment of ref.  \cite{LZGGZYGYP07}
a generalization of the schemes suggested in \cite{ZHWZ97} was used.
In our recent experiment \cite{RWZB09} we used a generalization of
the blueprint of \cite{ZW01}, and its realization \cite{GBEKW03}. We
obtained a six-photon invariant entangled state by pulse pumping
just {\em one crystal} and extracting the third order process. This
is done only via suitable filtering, and the interference is
observed behind four beamsplitters. The setup is strongly robust, as
it faces no alignment problems. The observed six-photon correlation
with  high fidelity agree with the ones of the theoretical
$\ket{\Psi_{6}^{-}}$.

In this paper present results of the invariance tests of the
experimental correlations attributable to  $\ket{\Psi_{6}^{-}}$.
This is done by sequence measurements of three mutually
complementary polarizations at all six detection stations (linear
vertical/horizontal, linear diagonal/antidiagonal, circular
right/left). The other interesting feature of the state is that it
reveals various interesting types of entanglement within the
subsystems. This is studied here  theoretically, and compared with
the data. Finally we present tests aimed at verification of the
entanglement of the obtained state. We use the toolbox of
entanglement witnesses provided in \cite{Toth-MATLAB}.

The state component corresponding to the emission in a PDC process
of six photons into two spatial modes in a  PDC process is
proportional to
\begin{equation}
(a_{0H}^{\dagger}b_{0V}^{\dagger}+e^{i\phi}a_{0V}^{\dagger}b_{0H}^{\dagger})^3\ket{0}.
\label{emission3}
\end{equation}
where $a_{0H}^{\dagger}$ ($b_{0V}^{\dagger}$) is the creation
operator for one horizontal (vertical) photon in mode $a_{0}$
($b_{0}$), and conversely; $C$ is a normalization constant, $\alpha$
is a function of pump power, non-linearity and length of the
crystal, $\phi$ is the phase difference between horizontal and
vertical polarizations due to birefringence in the crystal, and
$\ket{0}$ denotes the vacuum state. This is a good description of
the initial six-photon state, provided one collects the photons
under conditions that allow the indistinguishability between
separate two-photon emissions \cite{ZZW95}. A particle
interpretation of this term can be obtained through its expansion
\begin{align}
&(a_{0H}^{\dagger 3}b_{0V}^{\dagger 3}+3e^{i\phi}a_{0H}^{\dagger
2}b_{0V}^{\dagger 2}a_{0V}^{\dagger}b_{0H}^{\dagger}+ \nonumber \\
&+3e^{2i\phi}a_{0H}^{\dagger}b_{0V}^{\dagger}a_{0V}^{\dagger
2}b_{0H}^{\dagger 2}+e^{3i\phi}a_{0V}^{\dagger 3}b_{0H}^{\dagger
3})\ket{0}, \label{expansion}
\end{align}
and is given by the following superposition of photon number states:
\begin{align}
&\ket{3H_{a_{0}},3V_{b_{0}}}+e^{i\phi}\ket{2H_{a_{0}},1V_{a_{0}},2V_{b_{0}},1H_{b_{0}}}+
\nonumber \\
&e^{2i\phi}\ket{1H_{a_{0}},2V_{a_{0}},1V_{b_{0}},2H_{b_{0}}}+e^{3i\phi}\ket{3V_{a_{0}},3H_{b_{0}}},
\label{particle}
\end{align}
where e.g. $3H_{a_{0}}$ denotes three horizontally polarized photons
in mode $a_{0}$. The third order PDC is fundamentally and
intrinsically different than a product of three entangled pairs. Due
to the bosonic nature of photons the emissions of completely
indistinguishable photons are favored compared with the ones with
orthogonal polarization.

We report   experimental observations which are aimed at testing
whether the correlations produced in our setup are indeed
rotationally invariant. The  invariant six-qubit polarization
entangled state given by the following superposition of a six-qubit
Greenberger-Horne-Zeilinger (GHZ) state and two products of
three-qubit W states.
\begin{equation}
\ket{\Psi_{6}^{-}} = \frac{1}{\sqrt{2}}\ket{GHZ_{6}^{-}} +
\frac{1}{2}(\ket{\overline{W}_{3}}\ket{W_{3}}
-\ket{W_{3}}\ket{\overline{W}_{3}}). \label{state}
\end{equation}
The GHZ state  is here defined as
$\ket{GHZ_{6}^{-}}=\frac{1}{\sqrt{2}}(\ket{HHHVVV}-\ket{VVVHHH})$,
and the W-state is  defined as
$\ket{W_{3}}=\frac{1}{\sqrt{3}}(\ket{HHV}+\ket{HVH}+\ket{VHH})$.
$\ket{\overline{W}}$ is the spin-flipped  $\ket{W}$, and $H$ and $V$
denote horizontal and vertical polarization, respectively. This
state is obtained from the third order emission of the PDC process
eq. (\ref{emission3}) with the phase $\phi = \pi$. The emitted
photons are beam-split into six modes and one selects the terms with
one photon in each mode.

It is easy to see that if one moves into the spin description of the
polarization variables, the state is a singlet (total spin equal to
zero) of a composite system consisting of six spins $1/2$.


\begin{figure}
\includegraphics[width= \columnwidth]{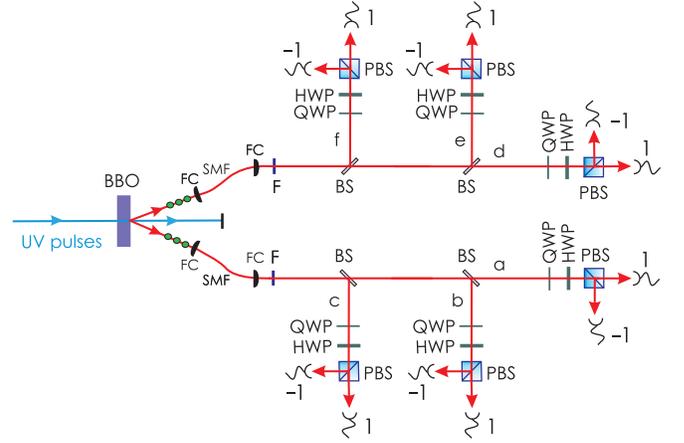}
\caption{\label{setup}Experimental setup for generating and
analyzing the six-photon polarization-entangled state.  The six
photons are created in third order PDC processes in a 2 mm thick BBO
pumped by UV pulses. The intersections of the two cones obtained in
non-collinear type-II PDC are coupled to single mode fibers (SMF)
wound in polarization controllers. Narrow band interference filters
(F) ($\Delta\lambda =3$ nm) serve to remove spectral
distinguishability. The coupled spatial modes are divided into three
modes each by 50\%-50\% beam splitters (BS). Each mode can be
analyzed  in arbitrary basis using half- and quarter wave plates
   (HWP and QWP) and a polarizing beam splitter (PBS). Simultaneous detection
   of six photons (two single photon detectors for each mode) are being recorded
   by a twelve channel coincidence counter.}
\end{figure}

In our experiment we use a frequency-doubled Ti:Sapphire laser ($80$
Mhz repetition rate, $140$ fs pulse length) yielding UV pulses with
a central wavelength at $390$ nm and an average power of $1300$ mW.
The pump beam is focused to a $160$ $\mu$m waist in a $2$ mm thick
BBO ($\beta$-barium borate) crystal. Half wave plates and two $1$ mm
thick BBO crystals are used for compensation of longitudinal and
transversal walk-offs. The third order emission of non-collinear
type-II PDC is then coupled to single mode fibers (SMF), defining
the two spatial modes at the crossings of the two frequency
degenerated down-conversion cones. Leaving the fibers the
down-conversion light passes narrow band  ($\Delta\lambda =3$ nm)
interference filters (F) and is split into six spatial modes $(a, b,
c, d, e, f)$ by ordinary $50\%-50\%$ beam splitters (BS), followed
by birefringent optics to compensate phase shifts in the BS's. Due
to the short pulses, narrow band filters, and single mode fibers the
down-converted photons are temporally, spectrally, and spatially
indistinguishable \cite{ZZW95}, see Fig.~\ref{setup}. The
polarization is being kept by passive fiber polarization
controllers. Polarization analysis is implemented by a half wave
plate (HWP), a quarter wave plate (QWP), and a polarizing beam
splitter (PBS) in each mode. The outputs of the PBS's are lead to
single photon silicon avalanche photo diodes (APD) through multi
mode fibers. The APD's electronic responses, following photo
detections, are being counted by a multi channel coincidence counter
with a $3.3$ ns time window. The coincidence counter registers any
coincidence event between the $12$ APD's as well as single detection
events.
\begin{figure}
\includegraphics[width= \columnwidth]{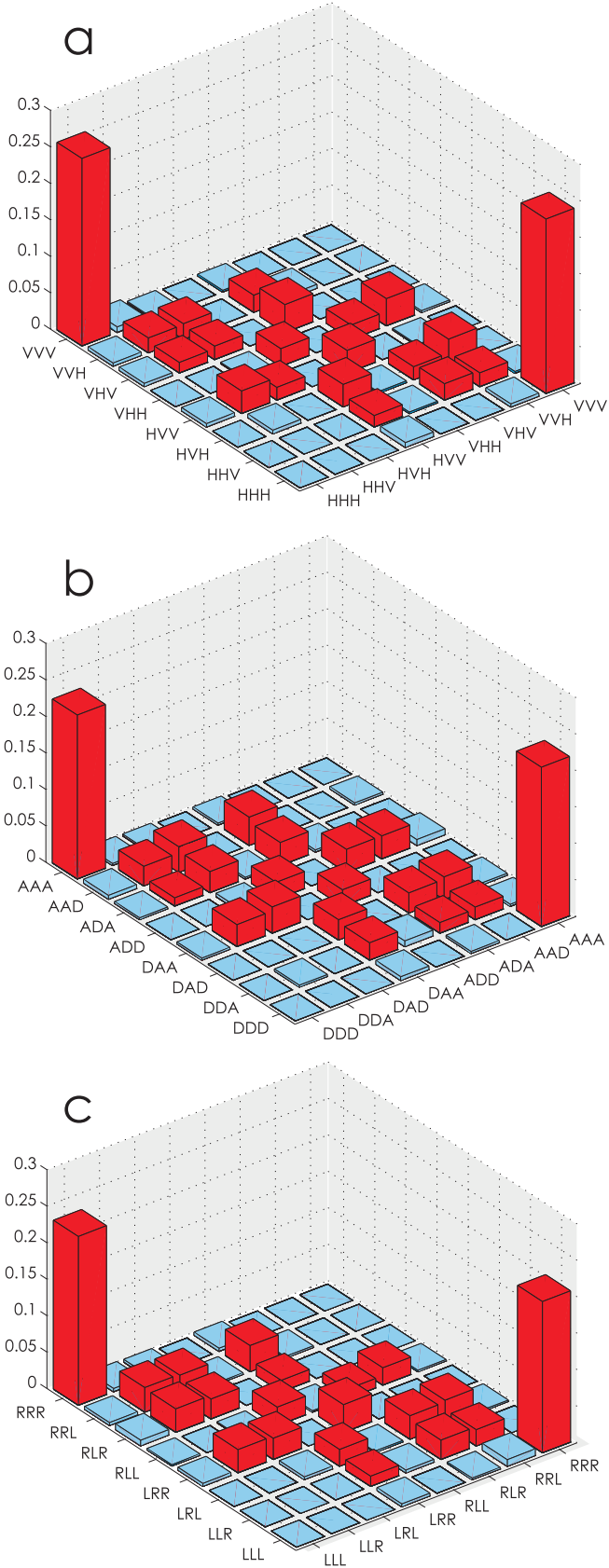}
\caption{\label{data}Experimental results of the six-photon
invariant state. Six-fold coincidence probabilities corresponding to
detections of one photon in each mode in $H/V$-basis (a),
$D/A$-basis (b), and $L/R$-basis (c).
  The values of the correlation functions are $-87.88\%\pm4.46\%$,
   $-87.59\%\pm4.97\%$, and $-86.79\%\pm4.27\%$ respectively.
   Comparing the three measurement results makes the invariance
   of our state obvious. For the pure $\ket{\Psi_{6}^{-}}$
   state the light blue bars would be zero and in our experiment
    their amplitudes are all in the order of the noise.
    The measurement time was about $140$ hours for each setting.
All data clearly shows
   the absolute value of amplitude and the coherence in the state.}
\end{figure}

Fig.~\ref{data}a shows the probabilities to obtain each of the $64$
possible sixfold coincidences with one photon detection in each
spatial mode, measuring all qubits in $\{\ket{H},\ket{V}\}$ basis.
The peaks are in very good agreement with theory:  half of the
detected sixfold coincidences are to be found as $HHHVVV$ and
$VVVHHH$, and the other half should be evenly distributed among the
remaining events with three $H$ and three $V$ detections. This is a
clear effect of the bosonic interference (stimulated emission) in
the BBO crystal giving higher probabilities for emission of
indistinguishable photons.

The six-photon state $\ket{\Psi_{6}^{-}}$ is invariant under
identical (unitary) transformations $U$ in each mode. Experimentally
this can be shown by  using  {\em identical} settings of all
polarization analyzers: no matter what the setting is,  the results
should be similar. Our results for measurements in
$\{\ket{D},\ket{A}\}$ (Diagonal/Antidiagonal, $\ket{D/A} = (\ket{H}
\pm \ket{V})/\sqrt{2}$) and $\{\ket{L},\ket{R}\}$ (Left/Right,
$\ket{L/R} = (\ket{H} \pm i\ket{V})/\sqrt{2}$) polarization bases
are presented in fig.~\ref{data}.b and \ref{data}.c. The invariance
of the probabilities with respect  the joint changes of the
measurement basis in all modes is clearly visible. We clearly
observe, in the results of these three different settings
measurements, the small and uniform noise contribution.

Another property of the $|\Psi_{6}^{-}\rangle$ is that it exhibit
perfect EPR correlations between measurement results in different
modes. We obtain the corrections
$\langle\Psi_{6}^{-}|\sigma_z^{\otimes 6}|\Psi_{6}^{-}\rangle = -
0.879 \pm 0.045$, $\langle\Psi_{6}^{-}|\sigma_x^{\otimes
6}|\Psi_{6}^{-}\rangle = - 0.876 \pm 0.050$, and
$\langle\Psi_{6}^{-}|\sigma_y^{\otimes 6}|\Psi_{6}^{-}\rangle = -
0.868 \pm 0.043$. which are close the theoretical value of $-1$.
From these results and the approximation that our noise to white
noise we have estimate the fidelity $F=
\langle\Psi_{6}^{-}|\rho_{exp}|\Psi_{6}^{-}\rangle =  0.876 \pm
0.045$ where $\rho_{exp}$ is the experimental the six-photon
density. The estimated fidelity clearly shows that the setup is able
to produce correlations due to six photon entangled states with
unprecedented precision (error rate below $6\%$).

Conditioning on a detection of  one photon in a specific state we
have also obtained four different five-photon   entangled
states. In the computational basis the projection of the last qubit
on $\ket{V}$ leads to
\begin{align}
_{f}\langle V \mid
\Psi_{6}^{-}\rangle=&\frac{1}{\sqrt{2}}\ket{HHHVV} \nonumber
\\ &-\frac{1}{\sqrt{3}}
\ket{W_{3}}\ket{\Psi_{2}^{+}}+\frac{1}{\sqrt{6}}\ket{\overline{W}_{3}}\ket{HH}.
\label{state5H}
\end{align}
A similar projection on $\ket{H}$ results in
\begin{align}
_{f}\langle H \mid \Psi_{6}^{-}\rangle
=&-\frac{1}{\sqrt{2}}\ket{VVVHH} \nonumber
\\ & +\frac{1}{\sqrt{3}}
\ket{\overline{W}_{3}}\ket{\Psi_{2}^{+}}-\frac{1}{\sqrt{6}}\ket{W_{3}}\ket{VV}.
\label{state5V}
\end{align}
We have also performed  such measurements related with the operator
$\sigma_z$ in the mode $b$, which
 has as its eigenstates
$\{\ket{H}_{b},\ket{V}_{b}\}$, while the other five photons are
measured in the $\{\ket{D},\ket{A}\}$ basis. The projection on
$\ket{H}_{b}$ gives
\begin{align}
_{b}\langle H \mid
\Psi_{6}^{-}\rangle=&\frac{1}{\sqrt{2}}\ket{GHZ_{5}^{-}}+\frac{1}{\sqrt{6}}
\big(\ket{\Psi_{2}^{+}} +\frac{1}{\sqrt{2}}\ket{AA}\big)\ket{W_{3}}
\nonumber \\ & -\frac{1}{\sqrt{6}}
\big(\ket{\Psi_{2}^{+}}+\frac{1}{\sqrt{2}}\ket{DD}\big)\ket{\overline{W}_{3}},
\label{state5D}
\end{align}
where
$\ket{GHZ_{5}^{-}}=\frac{1}{\sqrt{2}}(\ket{DDAAA}-\ket{AADDD})$ and
the projection on $\ket{V}_{b}$ gives
\begin{align}
_{b}\langle V \mid \Psi_{6}^{-}\rangle
=&\frac{1}{\sqrt{2}}\ket{GHZ_{5}^{+}}
-\frac{1}{\sqrt{6}}\big(\ket{\Psi_{2}^{+}}
-\frac{1}{\sqrt{2}}\ket{AA}\big)\ket{W_{3}} \nonumber \\ & -
\frac{1}{\sqrt{6}}\big(\ket{\Psi_{2}^{+}}-\frac{1}{\sqrt{2}}\ket{DD}\big)\ket{\overline{W}_{3}},
\label{state5A}
\end{align}
where
$\ket{GHZ_{5}^{+}}=\frac{1}{\sqrt{2}}(\ket{DDAAA}+\ket{AADDD}).$

Fig. 3 shows the results (obtained in the observation bases) for
these five photon conditional polarization states. In Fig. 3.a and
3.b, we clearly see the terms $\ket{VVVHH}$, and $\ket{HHHVV}$
respectively. The terms $\ket{DDAAA}$ and $\ket{AADDD}$ are evident
in both Fig. 3.c and 3.d. All these results are in agreement with
theoretical predictions.
\begin{figure}
\includegraphics[width= \columnwidth]{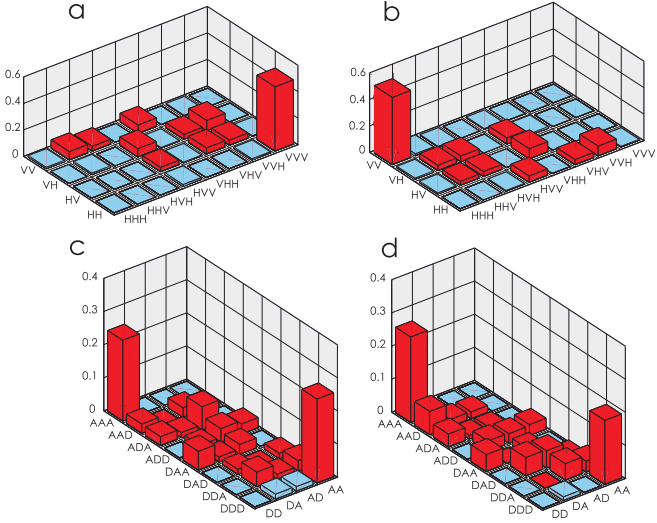}
\caption{\label{data5}Five-photon states from projective
measurements. Five-fold coincidence probabilities obtained through
the projection in $H/V$-basis of one photon. In (a) and (b) all the
qubits are measured in $H/V$-basis and the last qubit(in the mode f)
is projected onto H and V respectively. The results in (c) and (d)
correspond to projective measurements in $H/V$-basis on the qubit b
with the result $H$ and $V$ respectively, while the remaining five
photons are measured in $D/A$-basis.}
\end{figure}

$\ket{\Psi_{6}^{-}}$ is a  six-qubit entangled state, meaning that
each of its qubits is entangled with all the remaining ones. In
order to show that our experimental correlations reveal a
six qubit entanglement we use the entanglement witness method. An
entanglement witness is an observable yielding a negative value only
for entangled states, the most common being the maximum overlap
witness ($\WW_{max}$), which is the best witness with respect to
noise tolerance \cite{witness}.
The maximum overlap witness optimized for $\ket{\Psi_{6}^{-}}$ has
the form
\begin{equation}
\WW_{max}=\frac{2}{3}\eins^{\otimes6}-\ket{\Psi_{6}^{-}}\bra{\Psi_{6}^{-}},
\end{equation}
where the factor 2/3 is the maximum overlap of $\ket{\Psi_{6}^{-}}$
with any biseparable state. The witness detects   six-partite
entanglement with a noise tolerance around $34\%$, but it also
demands a large amount of measurement settings. Since it would be an
experimentally very demanding task to perform all these
measurements, we have developed a reduced witness that can be
implemented using only three measurement settings (this was done
using the tools provided by Toth \cite{TG05, Toth-MATLAB}). Our
reduced witness $\WW$, is given by
\begin{align}
\WW&=\frac{77}{288}\eins^{\otimes6}+\frac{1}{576}\sum_{i=x,y,z}\big(3\sigma_{i}^{\otimes2}
\eins^{\otimes4}+3\sigma_{i}\eins\sigma_{i}\eins^{\otimes3} \nonumber \\
&
+3\eins\sigma_{i}^{\otimes2}\eins^{\otimes3}+3\eins^{\otimes3}\sigma_{i}^{\otimes2}
\eins+5\sigma_{i}^{\otimes2}\eins\sigma_{i}^{\otimes2}\eins+5\sigma_{i}\eins\sigma_{i}^{\otimes3}\eins
 \nonumber \\
&
+5\eins\sigma_{i}^{\otimes4}\eins+3\eins^{\otimes3}\sigma_{i}\eins\sigma_{i}+5\sigma_{i}^{\otimes2}
\eins\sigma_{i}\eins\sigma_{i}+5\sigma_{i}\eins\sigma_{i}^{\otimes2}\eins\sigma_{i} \nonumber \\
&
+5\eins\sigma_{i}^{\otimes3}\eins\sigma_{i}+3\eins^{\otimes4}\sigma_{i}^{\otimes2}+5\sigma_{i}
^{\otimes2}\eins^{\otimes2}\sigma_{i}^{\otimes2}+5\sigma_{i}\eins\sigma_{i}\eins\sigma_{i}^{\otimes2}
 \nonumber \\
&
+5\eins\sigma_{i}^{\otimes2}\eins\sigma_{i}^{\otimes2}+9\eins^{\otimes6}-[\eins\leftrightarrow\sigma_{i}]
\big), \label{witness}
\end{align}
where $[\eins\leftrightarrow\sigma_{i}]$ denotes the same terms as
in the sum but with $\eins$ and $\sigma_{i}$ interchanged. This is
obtained from the maximum overlap witness as follows. First the
maximum overlap witness is decomposed into direct products of Pauli
and identity matrices, secondly only terms that are products of one
type of Pauli matrices and identity matrices are selected (all terms
that include products of at least two different Pauli matrices are
deleted, remaining only non-mixed terms), e.g.
$\sigma_{i}^{\otimes3}\eins\eins\sigma_{i}$, $i=x,y,z$. Finally, the
constant in front of $\eins^{\otimes6}$ in the first term of
eq.~(\ref{witness}) is chosen to be the smallest possible such that
all entangled states found by the reduced witness are also found by
the maximal overlap witness. Our reduced witness detects sixpartite
entanglement of $\ket{\Psi_{6}^{-}}$ with a noise tolerance of
$15\%$. The theoretical expectation value $\langle\WW\rangle=-1/18
\approx-0.056$ and our experimental result is
$\langle\WW\rangle=-0.023\pm0.012$, showing   entanglement with 2.0
standard deviations.

In summary, we have  experimentally
 tested the property of rotational invariance of the six-photon state produced by our setup.
The state is indeed entangled, and various different entangled
states can be obtained out of it with the use of projective
measurements of one of the qubits.
 We would like to note that the interference
contrast is high enough for our setup to be used in demonstrations
of various six-party  quantum informational applications (quantum
reduction of communication complexity of some joint computational
tasks,  secret sharing, {\em etc.}).

 {\bf Acknowledgements} This work was supported by
Swedish Research Council (VR).  M.\.{Z}. was supported by Wenner
Gren Foundations
and by the EU programme QAP (Qubit Applications, No. 015858).\\

\end{document}